# Vertical Thermospheric Density Profiles from EUV Solar Occultations made by PROBA2 LYRA for Solar Cycle 24

**E.M.B. Thiemann[1], M. Dominique[2], M.D. Pilinski[1], F.G. Eparvier[1]**

[1]Laboratory for Atmospheric and Space Physics, University of Colorado at Boulder, USA

[2]Royal Observatory of Belgium, Uccle, Belgium

Corresponding author: Edward Thiemann (thiemann@lasp.colorado.edu)

**Key Points:**

- $O + N_2$ density from 150-400 km is measured by broadband EUV solar occultations with systematic and random uncertainty less than 13% and 6%, respectively.
- Absolute density measurements are, on average, within 11% of NRLMSISE-00 predictions at 225 km prior to when instrument degradation becomes significant in 2013.
- The observed temperature variability at the terminators increases with increasing altitude, and EUV forcing becomes a secondary source of day-to-day variability above 300 km.

## Abstract

A new dataset of summed neutral $N_2$ and O number density profiles, spanning altitudes between 150 and 400 km, and observed during Northern Winter from 2010 through 2016 is presented. The neutral density profiles are derived from solar occultation measurements made by the 0.1-20 nm Zr channel on the LYRA instrument onboard PROBA2. The climatology derived from the vertical profiles is consistent with that predicted by the NRLMSISE-00 model, and the systematic error and random uncertainty of the measurements is less than 13% and 6%, respectively. The density profiles are used to characterize the response of thermospheric density to solar EUV irradiance variability. Peak correlation coefficients between neutral density and EUV irradiance occur near 300 km at the dusk terminator and 220 km at the dawn terminator. Density variability is higher at dawn than it is at dusk, and temperature variability increases with increasing altitude at both terminators.

## 1 Introduction

The measurement of neutral density profiles is valuable to addressing fundamental science and space weather interests, and in enhancing the understanding of the thermospheric state and its underlying processes. With the exception of limb scans made by the Global Ultraviolet Imager (GUVI) onboard the Thermosphere Ionosphere Mesosphere Energetics and Dynamics (TIMED) satellite, relatively few measurements of thermospheric density profiles above 120 km have been made over the past three decades. Model-measurement disagreement still persists in the most recent thermospheric models [e.g. Bruinsma and Forbes, 2007; Qian and Solomon, 2012], and more observations are needed, particularly below 300 km, to constrain and validate models. The altitude range between 100 and 300 km is particularly important because it is where temperature increases by a factor of four to seven as the thermosphere warms from the cold mesosphere to the hot exosphere. The dearth of modern observations is a result of the difficulty in observing this region of the thermosphere and has been termed the "Thermospheric Gap" [Oberheide et al., 2011]. Specifically, since 1993, measurements of thermospheric vertical density profiles above 125 km have been limited to those made by GUVI between 2002 and 2007 [Meier et al., 2015]. In lieu of direct measurements, global average vertical density profiles have been approximated from vertically distributed spacecraft accelerometers [Emmert, 2009]. This letter presents new thermospheric density measurements from 150 to 400 km made by extreme ultraviolet (EUV) solar occultations between 2010 and 2017.

Solar occultations have been used to probe Earth's thermospheric density in the past, with an emphasis on $O_2$. Aiken et al. [1993] report the most recent observations of $O_2$ density found by solar occultations, using wavelengths between 137 and 140 nm to measure $O_2$ density up to 220 km; and references therein describe preceding measurements of $O_2$ density by solar occultations. Prior observations of other species were made by Reid [1971], who used solar occultations in the 120-350 nm range made by OSO4 to estimate $N_2$ and O densities. Recently, Slemzin et al. [2016] used solar occultations from EUV imagers onboard the CORONAS and Project for Onboard Autonomy 2 (PROBA2) satellites to measure the signal extinction between 200 and 400 km, and compared the measured extinction with that predicted by two semi-empirical models without retrieving density profiles directly.

Aside from Slemzin et al. [2016], prior thermospheric solar occultation measurements were made using full disk irradiance measurements, where the primary challenge for thermospheric density retrievals is accounting for the spatial extent of the solar disk, which is comparable to the retrieved density scale-heights. At wavelengths shortward of 200 nm, the inhomogenous distribution of radiance over the solar disk further complicates density retrievals from UV solar occultations. Roble and Norton [1972] developed a method to account for these issues by using solar images and reference atmospheres in a forward model of the observed extinction at each tangent height ($h_t$), the height above the surface tangent to the line-of-sight.

In this paper, new 150 to 400 km thermospheric density profiles are presented that were observed over the winter hemisphere by the 1-20 nm Zr channel of the solar observing Large Yield Radiometer (LYRA) instrument [Hochedez et al., 2006; Dominque et al., 2013] on the European Space Agency (ESA) PROBA2 satellite [Santandrea et al., 2013] between 2010 and 2017. It is shown that full disk solar irradiance measurements from the Zr channel are suitable for measuring the summed O and $N_2$ number density at approximately 5 km resolution to within 15% absolute accuracy. Density profiles near solar minimum and maximum are compared, which show solar EUV forcing dominates the short and long term variability as expected. These new measurements are used to determine the sensitivities of scale height on solar EUV irradiance at both the dawn and dusk terminators.

We anticipate that these new vertical density profiles will be of interest to researchers studying the Earth's upper atmosphere as they provide new observations within the Thermospheric Gap that can be used to uncover new phenomena and to provide model constraints. These data will also be of interest to the operational space weather community because LYRA is currently providing its data with low latency in support of ESA space weather operations. As such, LYRA can provide near real-time observations of thermospheric density and temperature for atmospheric drag model assimilation. Furthermore, these data should be of interest to researchers of the Mars upper atmosphere because the Mars Atmosphere and Volatile Evolution (MAVEN) mission currently orbiting Mars includes a solar EUV Monitor (EUVM) [Eparvier et al., 2015] with a 17-22 nm channel capable of making similar observations at Mars.

## 2 Data and Methods

Solar occultation measurements are made using the LYRA instrument onboard PROBA2 launched by ESA in 2009. LYRA monitors the full-disk solar irradiance at high cadence (nominally 20 Hz, resampled to 1 Hz for occultations) in four broad channels: Lyman-alpha (120-123 nm), Herzberg (190-222 nm), Al (1-80 nm), and Zr (1-20 nm), and three redundant units: The primary unit, used for monitoring the Sun in a quasi-uninterrupted way; the back-up unit, used for special observation campaigns including occultations; and a third unit, reserved for calibration purposes.

PROBA2 flies in a retrograde sun-synchronous orbit (~700 km of altitude, 98° of inclination), which generates brief occultations for three months each year (from early November through early February) when the spacecraft enters eclipse at the dawn terminator, and exits it at the dusk terminator. At these moments, LYRA can observe the Sun through the Earth's atmosphere, and the detected signal is attenuated by an amount that depends on the altitude, the channel spectral response and the atmospheric composition. During eclipse season, the backup unit makes occultation observations during one orbit each day.

The spectral response of the instrument was measured prior to launch during calibration campaigns at the PTB/BESSY II synchrotron as reported in BenMoussa et al., [2009]. However, continuous solar exposure has caused severe degradation of the primary unit that significantly modifies the instrument spectral response [BenMoussa et al., 2013]. Simulations suggest that the degradation is from growth of a C layer on the optics. The backup unit also shows signs of degradation, but to a lesser extent. The backup unit underwent a marked increase in degradation caused by continuous exposure between 12 May 2012 and 30 May 2012, and degradation continued to increase at an elevated rate thereafter, as the rate of special campaigns increased. The Zr channel seems to be the least sensitive to degradation, although, the signal acquired by this channel has decreased by approximately 30% between January, 2010 and April, 2017. The reduced Zr channel degradation can be explained by the fact that this channel is mostly sensitive to short wavelengths (below 20 nm), where the absorption cross-section of the C contaminant is smaller. However, the presence of the C contaminant will bias the atmospheric densities deduced from the absorption by the Earth atmosphere and will be discussed further in Section 4.1.

The method of Roble and Norton [1972] is used to retrieve densities from the backup unit Zr channel measurements. While the high-level steps of the retrieval process are summarized here, this method is complex and the reader should refer to the original work for further detail. The fundamental measurement is the ratio of the measured intensity divided by the intensity at the top of the atmosphere, termed the Extinction Ratio (ER),

$$\boldsymbol{ER} = \boldsymbol{e}^{-\sum \boldsymbol{N_i \sigma_i}}, \qquad (1)$$

where $N_i$ and $\sigma_i$ are the column density and cross-section for the $i^{th}$ species.

The absorption cross-sections of O and $N_2$ are nearly identical over the Zr pass-band. Therefore, the Zr channel occultations cannot distinguish between the two species, and retrieved densities are the sum of the O and $N_2$ densities. The O cross-sections reported by Huebner et al. [1992] are used for both O and $N_2$. The assumption of equivalent O and $N_2$ cross-sections introduces a small systematic error which will be addressed in Section 3.

The column density ($N_{O+N2}(h_t)$) at a particular tangent height is found using a forward model of ER that accounts for the extinction of the spatially extended Sun through a model reference atmosphere. The reference atmosphere is assumed to be an exponentially decaying atmosphere of the form $\boldsymbol{n = n(z_0)exp((z - z_0)/H)}$ with a scale height ($H$). For the first iteration, $H$ is guessed, and then inferred from the retrieved density profiles in the subsequent iterations. In order to find $N_{O+N2}(h_t)$, the forward model solves for $\boldsymbol{n(z_0)}$ at each altitude while holding $H$ constant. Once $N_{O+N2}(h_t)$ is known, it is parametrically smoothed using an exponential smoothing function. This smoothed $N_{O+N2}$ profile is converted to a number density profile using an Abel transform under the assumption of spherical symmetry. When the retrieved scale heights above 250 km are within 5% of that in the reference atmosphere, the iteration ceases.

As an example, Figure 1 compares reference and retrieved density profiles for measurements made at the dawn terminator on 30 November 2010. The thin red lines show the reference atmospheres found while forward modeling the column densities for the first iteration. Each first-iteration reference atmosphere has the same initial scale height of 41.9 km. In general, the lower tangent height column density retrievals correspond with reference atmospheres closer to the right of those shown. The resulting column density profile (not shown) is Abel transformed, and the resulting number density profile is shown with the thick red dashed line. A power law fit to the retrieved density above 250 km is used to find *H* for the next iteration and, in this example, is found to be equal to 37.802 km. The reference atmospheres for the second iteration are shown with thin blue lines. The resulting (final) density retrieval is shown with the thick solid black line (this line appears dashed with the dashed-red line overlaid), and has a corresponding scale height above 250 km equal to 37.797 km, well within the 5% of the preceding reference atmosphere scale-height. The initial and final density profiles (red-dashed and black-solid lines) are very similar, differing by a maximum of 0.9% at 140 km, becoming in better agreement at intermediate altitudes, and diverging slightly at high altitudes, with a 0.35% difference at 300 km.

For comparison, a predicted profile from the NRLMSISE-00 model [Picone et al., 2002] is shown with a black dashed-dotted line. The NRLMSISE-00 model is a semi-empirical model of thermosphere neutral composition and density, and predicts climatology relatively accurately, while predicting day-to-day variability less accurately. The NRLMSISE-00 predictions are derived from observations, and provide a convenient means to compare new and past measurements. O and $N_2$ densities between 140 and 400 km are calibrated against in-situ mass-spectrometer and accelerometer measurements made in the 1970s and 1980s. Temperatures from 140-220 km have been augmented to agree with measurements of $O_2$ made by Aiken et al. [1993]. Comparing the final density retrieval with the predictions from NRLMSISE-00 shows consistency between model and measurement until just below the change in scale height, highlighting the limitations of using a single scale-height reference atmosphere.

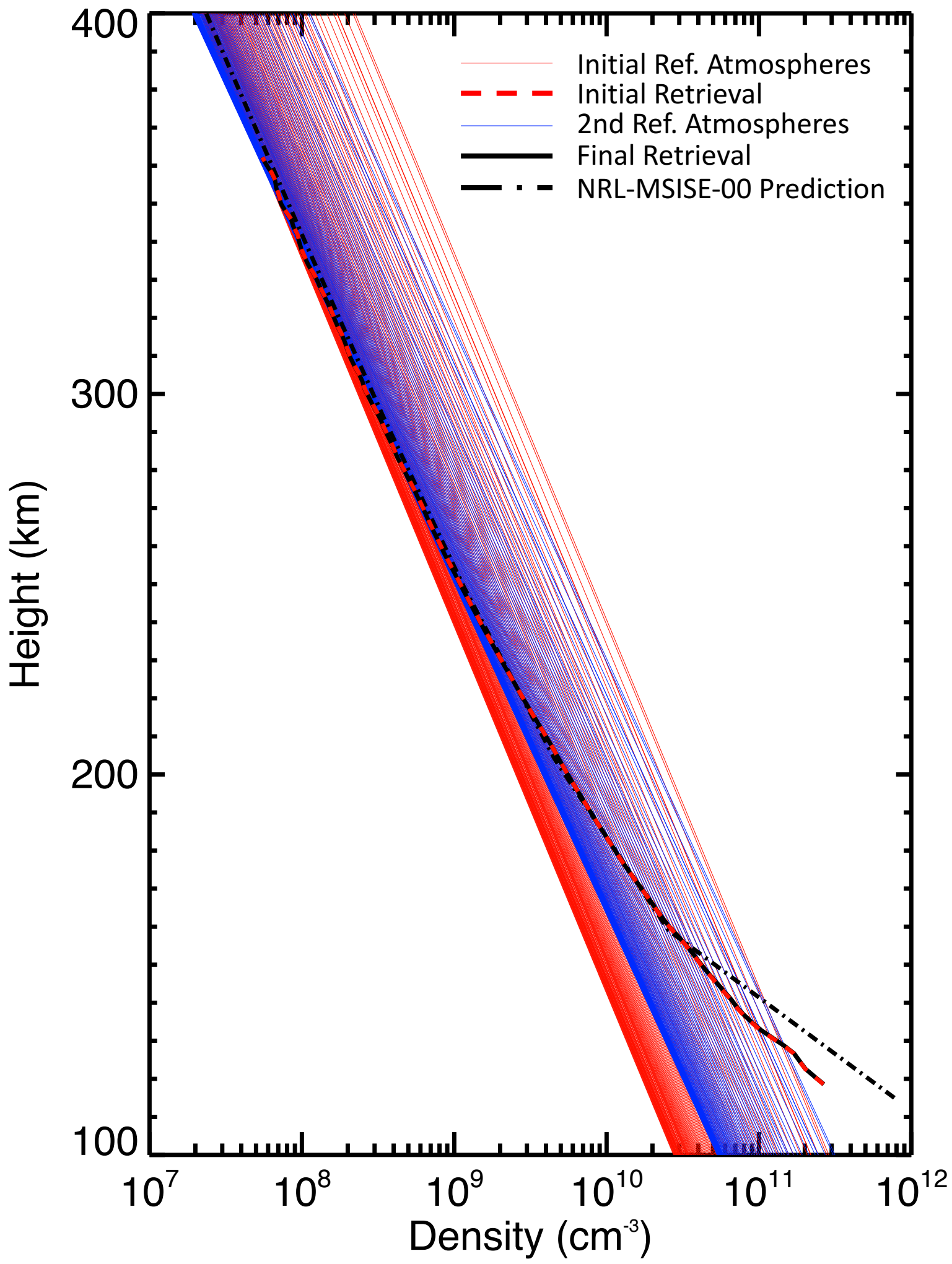

*Figure 1. Reference and retrieved atmospheres found during the 30 November 2010 dawn terminator density measurements. See Section 2 for details.*

A solar spectrum must be convolved with the (broadband) instrument response function to forward model equation (1). The retrieval method uses daily average solar spectra from the Flare Irradiance Spectral Model (FISM) originally developed by Chamberlin et al. [2007] and later updated by Thiemann et al. [2017]. This latest version of FISM has 0.1 nm resolution between 6 and 106 nm and 1 nm resolution otherwise. The ER is independent of absolute irradiance, but does depend on the relative irradiance calibration. As such, the relative uncertainty must be considered when propagating uncertainties. For the wavelengths in the Zr bandpass, the relative uncertainties are approximately 15% and 5%, shortward and longward of 6 nm, respectively.

The radiance distribution at 17.4 nm is representative of that in the Zr bandpass. Therefore, the spatial distribution of radiance over the solar disk in the Zr bandpass is determined from solar images taken by the 17.4 nm PROBA2 Sun Watcher using Active Pixel system detector and image processing (SWAP) telescope [Berghmans et al., 2006], which images the Sun 1-2 times per minute. The SWAP image taken nearest in time prior to the start (end) of a dawn (dusk) terminator occultation is used.

## 3 Measurement Uncertainty and Error

Numerical modeling, described below, is used to quantify uncertainty resulting from the cross-section assumptions, solar spectra, solar variability, and measurement noise. The instrument response function uncertainty is neglected because it is currently ill defined due to instrument degradation discussed in Section 2, but is negligible early in the mission when degradation is insignificant [BenMoussa et al., 2013]. As such, the uncertainty analysis is applicable for measurements prior to November, 2012, when degradation became apparent. The effect of the response function degradation is treated separately in Section 4.1.

Figure 2a shows curves for the $N_2$ and O cross-sections over the Zr channel bandpass in blue and red, respectively. The over-plotted black curve is the channel "Solar Sensitivity". This is the product of typical solar spectral irradiance on the detector and the Zr response function, which yields the photocurrent as a function of wavelength. Although the O and $N_2$ cross-sections are very similar at the dominant wavelengths, the differences add some error to the retrievals that is quantified as follows: First, the cross-sections are used to calculate the fractional error introduced on $ER$ ($\Delta ER(N_{N2})$) as a function of $N_2$ column density ($N_{N2}$) from using the O cross-section for $N_2$. From Equation (1), $\Delta ER(N_{N2})$ propagates to the fractional error on $N_{N2}$ ($\Delta N_{N2}$) as

$$\Delta \boldsymbol{N_{N2}} = \Delta \boldsymbol{ER}/\boldsymbol{ln}(\boldsymbol{ER}). \qquad (2)$$

For a given density retrieval, the NRL-MSISE-00 model is used to estimate $N_{N2}$ (and, in turn, $\Delta N_{N2}$). $\Delta N_{N2}$ is then propagated through the Abel transform to estimate, $\Delta n_{SYS,\sigma}$, the systematic error of the density retrievals resulting from using the O cross-section for $N_2$. The dotted curve in Figure 2b shows $\Delta n_{SYS,\sigma}$ for 14-Jan-2014, which is representative of solar moderate conditions.

Monte-Carlo simulation is used to estimate the uncertainty from solar spectra, solar variability and measurement noise. Ground-truth atmospheres consisting of $N_2$ and O for the times and positions corresponding with the LYRA measurements are computed from NRLMSISE-00, from which the ER is synthesized; random error representative of solar spectrum uncertainty is added to the solar spectrum used in the ER synthesis. Randomly generated measurement noise typical of the backup unit Zr channel is added, as well as solar variability typical of that observed over occultation measurement.

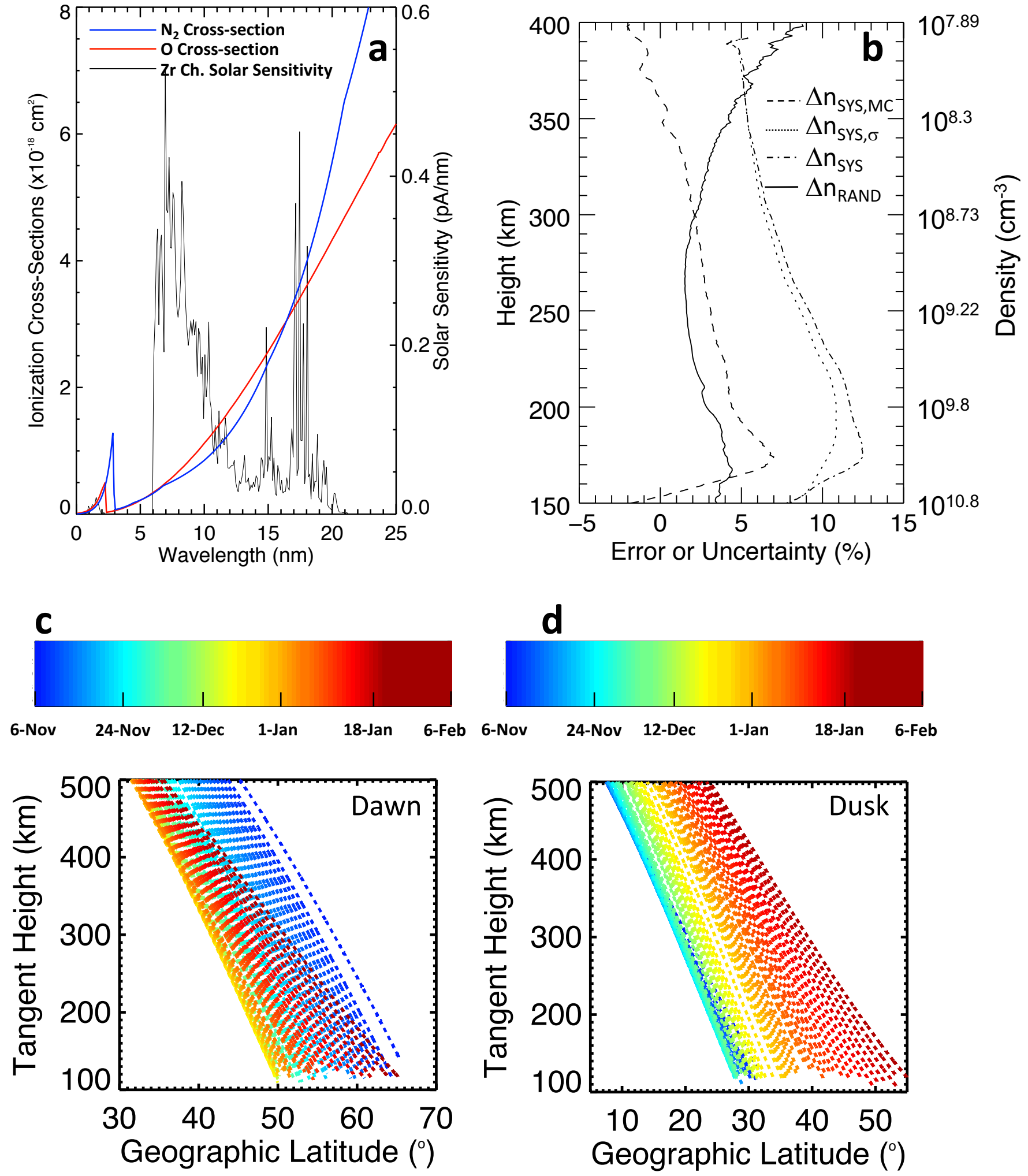

*Figure 2. Sources and components of measurement error and uncertainty a. $N_2$ (blue) and O (red) cross-sections across the Zr channel bandpass over plotted with typical solar spectral sensitivity in black. b. Modeled random uncertainties and systematic errors versus altitude (left axis) and density (right axis). See text for a definition of variables in the legend. c. Latitudes for tangent heights versus time at the dawn terminator. d. Same as c. but for the dusk terminator.*

The solar variability is quantified by first finding the change in solar irradiance over a time interval equal to the occultation measurement time (approximately 2 minutes) immediately before and after each occultation measurement. This is done for all occultation orbits, and the mean of the absolute value of this change is considered the typical variability. Note that orbits with solar flares are readily identified by significant enhancements in the Zr pass band and are discarded.

Figure 2b uses dashed and solid curves to show the systematic errors, $\Delta n_{SYS,MC}$, and random uncertainties, $\Delta n_{RAND}$, as a function of altitude determined by comparing the retrieved and ground-truth densities from 150 simulations, the corresponding average density is shown on the right vertical axis. $\Delta n_{RAND}$ values are near 3% for most altitudes, but become larger at both the top and bottom of the retrieved atmosphere. This is expected because the sample-to-sample change in the ER is smaller at these altitudes, resulting in a greater sensitivity to random sources of error. $\Delta n_{SYS,MC}$ values indicate that the density retrievals are typically 1-2% too tenuous at high altitudes and 5-7% too dense at low altitudes. This low altitude bias is expected because the reference atmosphere used in the retrieval only assumes a single scale height, whereas in reality, at lower altitudes, where $N_2$ becomes the major species, the scale height decreases by approximately one-half. The retrieval algorithm compensates for this by adding more density. The dashed-dot curve in Figure 2b shows the combined systematic uncertainty, $\Delta n_{SYS} = \Delta n_{SYS,\sigma} + \Delta n_{SYS,MC}$.

The latitude corresponding with the density profile geographic locations vary strongly with altitude due to the highly inclined spacecraft orbit. Figures 2c and 2d show the latitudes as a function of $h_t$ for the dawn and dusk terminator, respectively, where color indicates time. For example, at the dawn terminator for a single vertical profile, the latitudes range from $45^o$ to $65^o$ in early November, and $30^o$ to $50^o$ near early January. Since vertical density profiles are typically assumed to be in hydrostatic equilibrium when deriving a temperature from them, horizontal density gradients will increase the error on derived temperatures. One important consequence is that temperature profiles will converge on a constant exospheric value at approximately 10% higher altitudes.

## 4 Results

Sample density profiles near solar minimum and maximum are shown in Figure 3. Panels a and c correspond with dawn and dusk densities versus altitude, respectively, for the 2010-2011 occultation campaign, which is during minimum-to-moderate solar conditions. Panels d and f are similar, but for the 2013-2014 campaign occurring near solar moderate-to-maximum conditions. Although measurements are made daily, some days are omitted because the retrieval program failed to converge on a solution for these days. Less frequently, orbits are omitted because of solar flares or abnormal noise on the measurements. Panel g shows sample vertical profiles for 7-December from both campaigns. Panels b and e show the integrated 0.5 to 102.8 EUV flux from the FISM model. One difference between the two campaigns is the significantly higher density during higher solar conditions. Comparing the EUV fluxes with the corresponding density curves indicates that the observed thermospheric variability is tightly coupled with solar variability. These findings are not surprising and are consistent with past observations, and will be explored further in Section 4.2. Also, comparing the dawn vs dusk densities for the 2010-2011 (2013-2014) campaign, the densities are comparable near 250 km, but substantially more tenuous at 350 (400) km, indicative of a cooler dawn thermosphere.

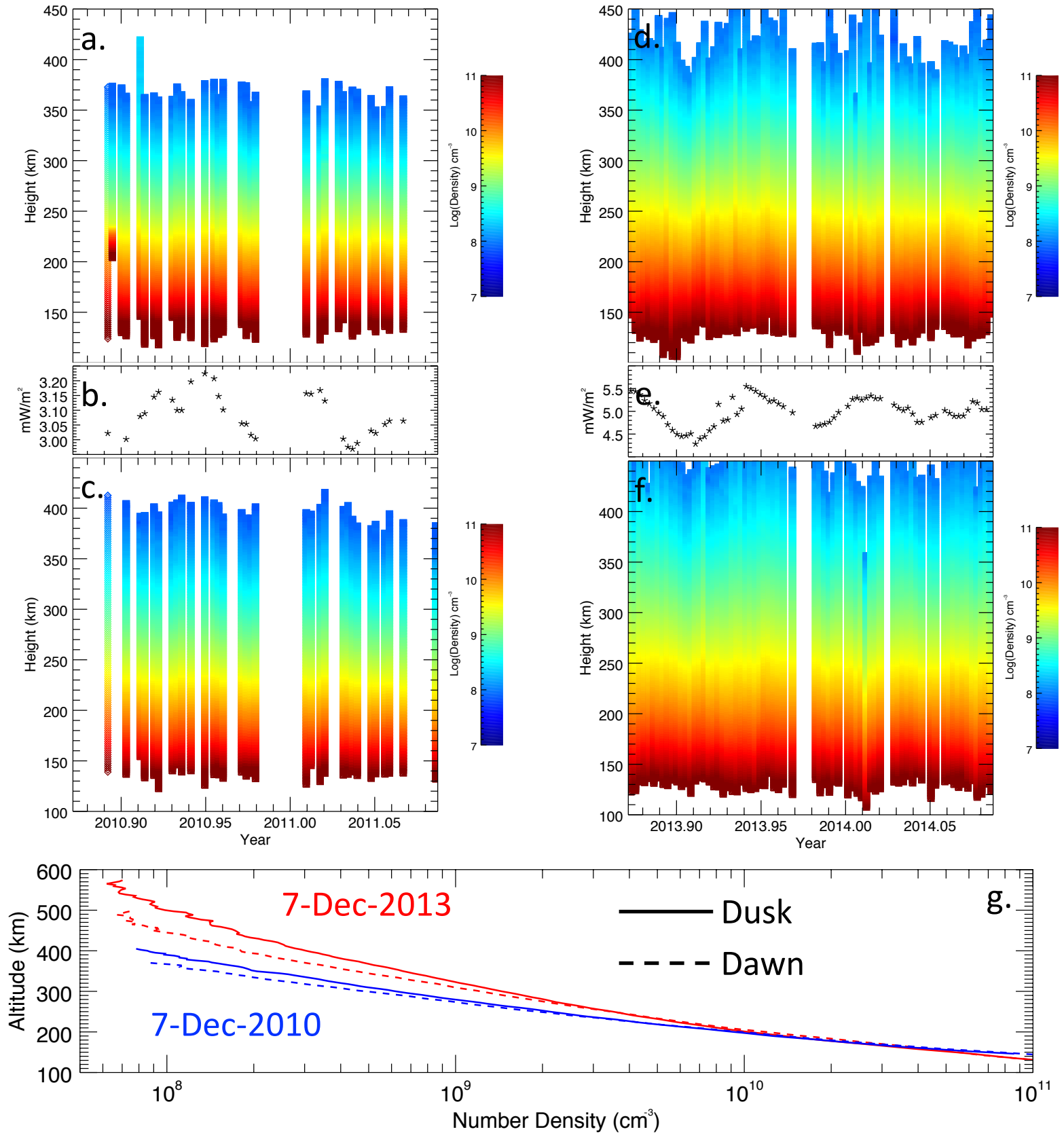

*Figure 3. Thermosphere O + $N_2$ densities for two campaigns with differing degrees of solar activity. Panels a and c are the dawn and dusk terminator densities, respectively, for the 2010-2011 occultation campaign, with altitude on the vertical axis, time on the horizontal axis and the color scale corresponding with logarithmic density. Panel b is the corresponding integrated 0 - 102.8 nm EUV irradiance. Panels c-f are the same as a-c but for the 2013-2014 occultation campaign. Panel g compares density profiles at different terminators and different stages of the solar cycle.*

Figure 4 shows time-series of densities from the 2013-2014 occultation campaign at 225 km (Panels e and f), 275 km (Panels c and d) and 325 km (Panels a and b) for both terminators. All panels show a local maximum near 2013.95 that corresponds with a local EUV maximum near the same time period, as seen in Figure 3e. However, this maximum is more structured at the dusk terminators (Panels b, d and f) than at the dawn terminators (Panels a, c and e). This reflects the general trend seen at all three altitudes of the dawn terminator being more variable than the dusk terminator. At the dusk terminator, the local maximum near 2013.95 becomes more pronounced with increasing altitude, with a 23% enhancement at 225 km and a 44% enhancement at 325 km. This indicates that the atmosphere is expanding as a result of increased EUV heating.

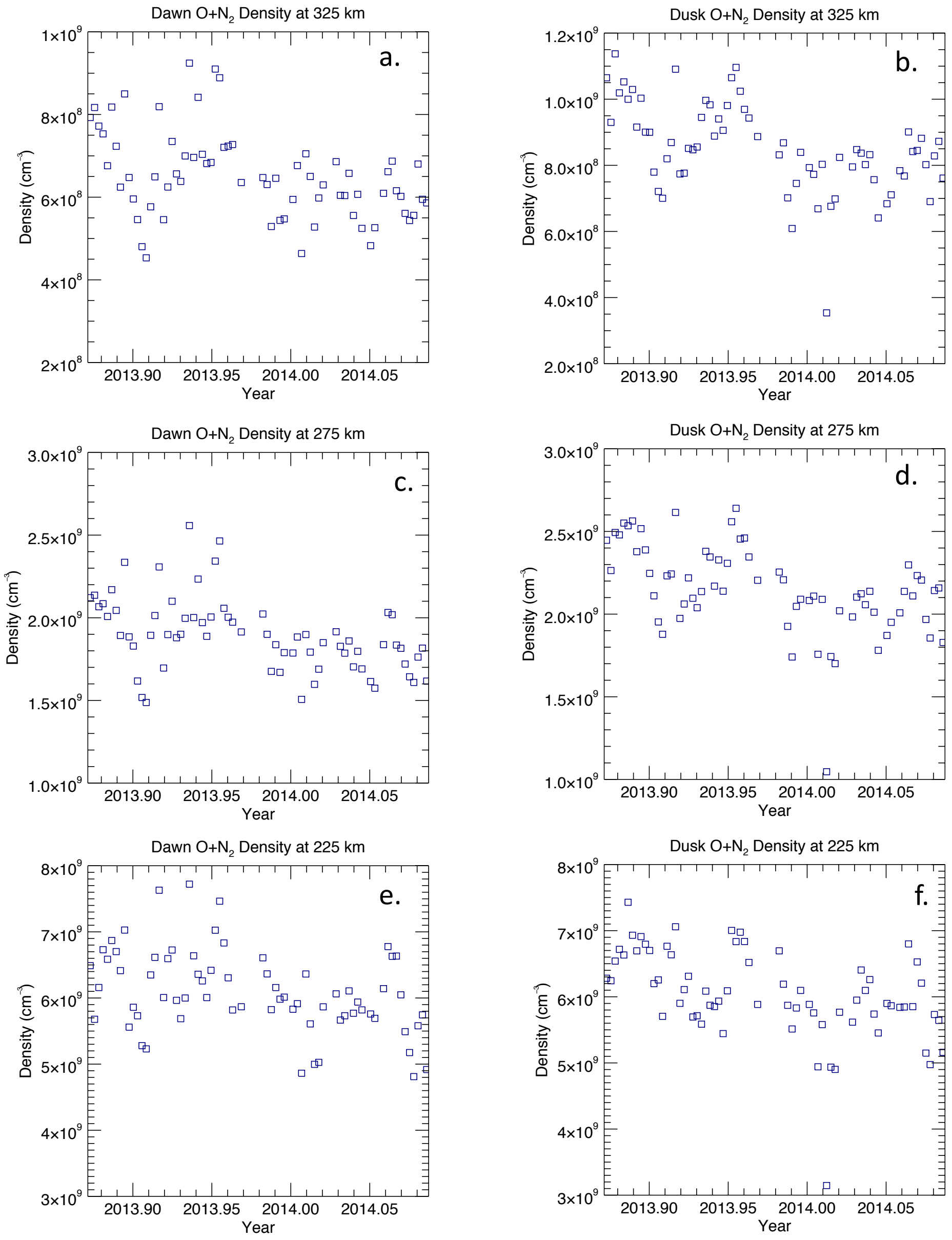

*Figure 4. Densities at fixed altitudes from the 2013-2014 occultation campaign. Dawn and dusk densities at 225 km (Panels e and f), 275 km (Panels c and d), and 325 km (Panels e and f) are shown.*

### 4.1 Comparison with the NRLMSISE-00 Model.

For this study, NRLMSISE-00 is used to generate $N_2$+O densities for each observed occultation profile. The ratio of the LYRA measurements to NRLMSISE-00 predictions is used to compare the profiles. Table 1 shows the campaign mean and standard deviation of this ratio at various altitudes for both terminators.

The LYRA measurement uncertainties are expected to be lowest for the 2010-2011 campaign, when the instrument response function had undergone the least degradation. For this campaign, the mean ratios are within 5% of unity at all altitudes except the 375 km dawn terminator value, which is a clear outlier. From Figure 2a, it is seen that the 375 km densities are near the measurement limits during the 2010-2011 campaign, hence, this value can be neglected. The dusk terminator mean ratios show an increasing trend with decreasing altitude, consistent with the modeled $\Delta n_{SYS}$ discussed in Section 3. However, this trend is not evident at the dawn terminator aside from a relative enhancement at 175 km. The model-measurement standard deviation is a combination of random measurement noise and model error. Below 325 km on the dawn terminator and 325 km on the dusk terminator, the standard deviations range from 0.08 to 0.11. This is significantly less than the model-measurement standard deviation of 0.17 reported by Picone et al. [2002] for the NRLMSISE-00 input measurements, indicating that the LYRA random measurement noise is smaller than that of the measurements which NRLMSISE-00 is based on (assuming the model error is constant).

*Table 1. Comparison of LYRA measured and NRLMSISE-00 modeled $N_2$+O densities at different altitudes and campaigns.*

| | | | Season | | | | | | |
|---|---|---|---|---|---|---|---|---|---|
| Side | Altitude | Moment | 2010-2011 | 2011-2012 | 2012-2013 | 2013-2014 | 2014-2015 | 2015-2016 | 2016-2017 |
| Dawn Terminator | 175 | Mean | 1.03 | 1.01 | 0.87 | 0.72 | 0.74 | 0.79 | 0.80 |
| | | StdDev | 0.09 | 0.13 | 0.11 | 0.09 | 0.12 | 0.10 | 0.07 |
| | 225 | Mean | 1.00 | 1.00 | 0.93 | 0.75 | 0.75 | 0.80 | 0.78 |
| | | StdDev | 0.11 | 0.08 | 0.08 | 0.08 | 0.06 | 0.07 | 0.06 |
| | 275 | Mean | 0.99 | 1.01 | 0.98 | 0.78 | 0.79 | 0.83 | 0.76 |
| | | StdDev | 0.11 | 0.10 | 0.10 | 0.09 | 0.09 | 0.10 | 0.08 |
| | 325 | Mean | 1.01 | 1.04 | 1.00 | 0.79 | 0.83 | 0.85 | 0.78 |
| | | StdDev | 0.20 | 0.14 | 0.16 | 0.11 | 0.11 | 0.15 | 0.13 |
| | 375 | Mean | 1.53 | 1.08 | 1.08 | 0.82 | 0.87 | 1.00 | 1.78 |
| | | StdDev | 0.61 | 0.17 | 0.22 | 0.15 | 0.16 | 0.24 | 0.56 |
| Dusk Terminator | 175 | Mean | 1.04 | 0.99 | 0.83 | 0.69 | 0.69 | 0.77 | 0.83 |
| | | StdDev | 0.08 | 0.10 | 0.12 | 0.10 | 0.09 | 0.09 | 0.06 |
| | 225 | Mean | 1.00 | 1.02 | 0.89 | 0.72 | 0.69 | 0.79 | 0.78 |
| | | StdDev | 0.08 | 0.08 | 0.09 | 0.08 | 0.07 | 0.06 | 0.07 |
| | 275 | Mean | 0.97 | 1.02 | 0.92 | 0.73 | 0.72 | 0.81 | 0.76 |
| | | StdDev | 0.08 | 0.08 | 0.09 | 0.09 | 0.07 | 0.06 | 0.07 |
| | 325 | Mean | 0.93 | 1.03 | 0.93 | 0.73 | 0.73 | 0.81 | 0.72 |
| | | StdDev | 0.10 | 0.10 | 0.09 | 0.10 | 0.09 | 0.07 | 0.11 |
| | 375 | Mean | 0.96 | 1.07 | 0.94 | 0.72 | 0.74 | 0.80 | 0.82 |
| | | StdDev | 0.34 | 0.13 | 0.13 | 0.13 | 0.12 | 0.08 | 0.14 |
| MSIS Correction Factors | | Dawn | 1.00 | 1.00 | 1.07 | 1.33 | 1.30 | 1.24 | 1.29 |
| | | Dusk | 1.00 | 0.98 | 1.11 | 1.38 | 1.40 | 1.25 | 1.28 |

The LYRA-to-NRLMSISE-00 comparisons are used to quantify the impact of instrument degradation and correct for it. The first two campaigns show no significant difference between the mean ratios. However, there is an approximate 10-15% drop between the 2011-2012 and 2012-2013 campaigns, and an additional 10-15% drop preceding the following campaign. The timing of this decrease corresponds with degradation resulting from the commanding anomaly and the subsequent special observing campaigns discussed in Section 2. However, this decrease may not all be from degradation because NRLMSISE-00 is known to overestimate density with increasing solar activity by comparable magnitudes [Picone et al., 2002]. The fact that the effect of degradation on the density retrievals is relatively altitude independent suggests the contaminant causing the degradation has an optical cross-section comparable to O, and supports C as being a likely candidate contaminant as discussed in Section 2.

To correct for this degradation, the LYRA to NRLMSISE-00 ratios are used to scale the densities to the 2010-2011 values as follows: For each campaign, the mean of the ratio mean between 175 and 325 km ($r_s$) is first found. Next, the densities occurring during the $s^{th}$ campaign are scaled by $r_{2010\text{-}2011}/r_s$. These correction factors are shown in the bottom two lines of Table 1 for the dawn and dusk terminators.

It should be emphasized that by using NRLMSISE-00 to derive the degradation factor, the error inherent in the NRLMSISE-00 predictions will propagate to the LYRA measurements. Recent accelerometer measurements have quantified the NRLMSISE-00 error near 270 km. Bruinsma et al. [2017] compared measurements from the precision accelerometer onboard the Gravity field and steady-state Ocean Circulation Explorer (GOCE) with NRLMSISE-00 density predictions at the GOCE orbit altitude of approximately 270 km, and showed the GOCE measurement to NRLMSISE-00 model ratio to be approximately 0.9 in January 2010. This ratio increased, apparently with solar activity, to approximately 1.1 in mid-2012, prior to when the GOCE altitude began to decline. As such, the correction factors in Table 1 may need to increase a further ~20% between 2010 and 2012 to be in better agreement with the GOCE measured densities at 270 km. A detailed day-to-day direct comparison of LYRA and GOCE measurements is necessary to quantify their interrelationship more definitively.

### 4.2 Dependence of Density on EUV Irradiance versus Altitude

The dependence of neutral density on solar EUV irradiance at various altitudes is a convenient metric for evaluating and constraining thermospheric models. To this end, the LYRA vertical number densities are used to quantify the dependence of neutral density on solar EUV irradiance. The integrated 0.5 to 102.8 nm irradiance is used as a proxy for solar EUV energy input into the thermosphere, where the longward wavelength cutoff corresponds with the ionization threshold of $O_2$, which has the lowest ionization potential of the major thermospheric neutral species.

The Pearson correlation coefficient quantifying the correlation of neutral density with EUV irradiance for both terminators is shown in Figure 5a as a function of altitude. Correlation coefficients at the dusk terminator are higher as is expected given the density peaks due to solar forcing near 15:00 Local Time. The dusk terminator correlation coefficients reach a maximum near 250 km, whereas the dawn terminator values peak much lower, near 220 km. Correlation coefficients at both terminators show a marked decrease with increasing altitude beginning at 300 km. The decreasing correlation coefficient is a result of increasing variability of non-solar origin. Figure 2b shows that the random uncertainty (i.e. measurement noise) increases above 300 km, as the extinction ratio approaches 1 (the average extinction ratio for the LYRA dataset is equal to 0.95 at 300 km), but this modest increase in random uncertainty is insufficient to account for the total increase in variability with altitude indicated by the decrease in correlation with altitude. As such, these results indicate that the thermospheric density at the terminator becomes increasingly variable with altitude above approximately 300 km; this will be considered further towards the end of this section.

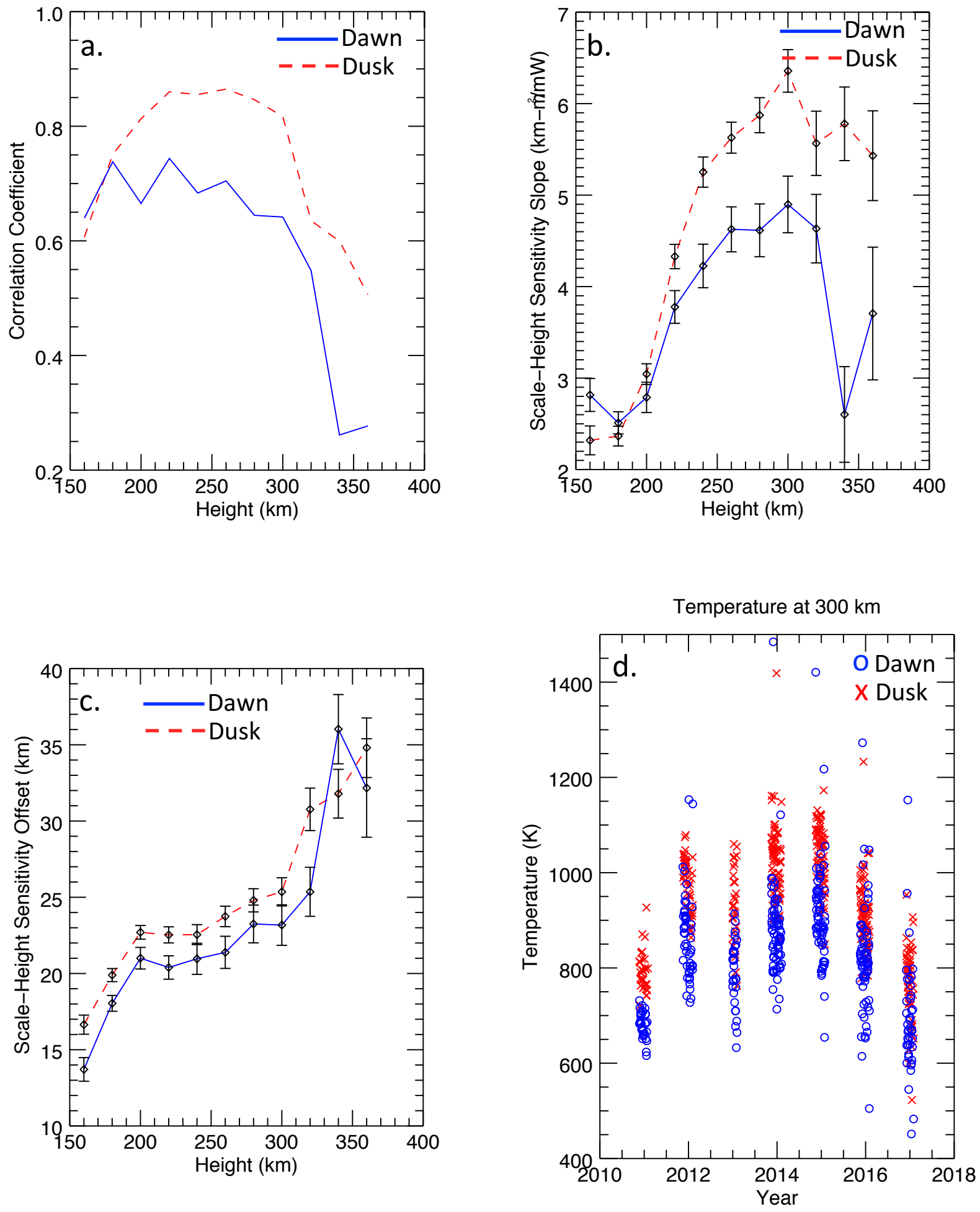

*Figure 5. Scale-height and temperature dependence on EUV irradiance. a. Pearson correlation coefficients between integrated ionizing EUV irradiance and density. b. and c. Slopes and offsets of linear fits between scale height and EUV forcing, providing a measure of the scale-height sensitivity on EUV forcing. d. Temperature at 300 km versus time at both terminators.*

The sensitivity of scale-height to EUV irradiance variability is calculated by finding first order, least-squares fits between the local scale-heights and EUV irradiance at 20 km intervals. The resulting slopes and offsets versus altitude are shown in Figures 5b and 5c, respectively. The scale-height, H, is found by fitting a power law of the form, $n = n(z_o) \cdot exp(-(z - z_o)/H)$ to the densities from the indicated altitude to 20 km above it. The error-bars shown correspond with the 1-sigma uncertainty, assuming a first-order linear equation is the correct model. This assumption likely breaks down as the correlation decreases, resulting in an underestimation of the uncertainty for the cases that show a low correlation between scale height and EUV irradiance.

The scale-height sensitivity can be converted into temperature sensitivity if the relative abundance of $N_2$ and O are known. At both terminators, the sensitivity slope peaks near 300 km. Since O abundances are typically 10X that of $N_2$ at this altitude, these data suggest that the O temperature sensitivity peaks at 300 km. Constant scale-height is reached near 325 km (higher than the nominal exospheric temperature height because the density profiles span a range of latitudes) and the temperature sensitivities at the dusk terminator become approximately constant above 325 km as expected. The dawn terminator values above 325 km in both Panels b and c are likely the result of spurious linear fits given the small values of the corresponding correlation coefficients. Lower in the atmosphere, the general decrease in scale-height sensitivity with decreasing altitude must in large part be due to $N_2$ becoming the major species with decreasing altitude rather than a decrease in temperature sensitivity at lower altitudes, which can only be investigated by accounting for the relative abundances.

The results in Figures 5a and 5b should be considered in the context of the re-analysis of neutral EUV heating efficiencies by Richards [2012], which shows that the inclusion of thermal electron heating broadens the neutral heating efficiency altitude profile and increases the heating efficiency at high altitudes, resulting in an approximate doubling of the neutral heating efficiency at 300 km. However, thermal electron heating is highly variable, and this is the likely explanation of the decreasing correlation coefficient above 300 km shown in 5a. Richards [2012] also shows that the heating per neutral particle reaches a constant value near 300 km, which may explain the asymptotic increase of the scale-height sensitivity to a near-constant value near 300 km at the dusk terminator.

Figure 5d shows the thermosphere temperature at 300 km at both terminators for the LYRA measurements over solar cycle 24. The dusk temperatures tend to be approximately 100 K warmer than the dawn temperatures, and the temperature increases with increasing solar activity (solar maximum occurred in mid-2013.)

## 5. Summary and Conclusions

Broadband solar EUV measurements between 5 and 20 nm are used to retrieve neutral densities from 150 to 400 km using the method of Roble and Norton [1972], covering a region of the atmosphere that has been historically difficult to measure. Systematic and random uncertainties are below 15%, and the absolute densities are within 5% of predictions by NRLMSISE-00. Above 225 km, the correlation between density and solar EUV irradiance increases with altitude at the dusk terminator, but decreases with altitude at the dawn terminator, until 300 km at which point the correlation decreases rapidly.

These LYRA retrievals provide 7 northern-winter campaigns of new vertical density profiles. The low random uncertainties make these data suitable for characterizing thermospheric variability due to space weather events such as solar flares and coronal mass ejections, as well as non-migrating tides. Because LYRA is currently used by ESA for space weather operations, it is possible for LYRA density measurements to be made rapidly publically available to provide near real time measurements of the thermosphere state for model assimilation.

Because modern EUV photometers have very low noise levels and make high cadence measurements, they are appealing for solar occultation measurements. The methods presented here are directly applicable to the 17-22 nm channel of the EUV Monitor onboard MAVEN for retrieving thermospheric $CO_2$ densities at Mars from solar occultations. The small size of an EUV photometer makes it a relatively low cost option for characterizing thermosphere density as part of research missions at Earth and other planets, or for routine monitoring as part of operational space weather missions.

**Acknowledgments, Samples, and Data**

LYRA irradiances are available on the web at http://proba2.oma.be/lyra/data/bsd/. LYRA density retrievals are available on the web at http://proba2.oma.be/lyra/data/EarthAtmosphere/.

SWAP images are available on the web at http://proba2.oma.be/swap/data/bsd/

Photo-ionization cross-sections were downloaded from the PHoto Ionization/Dissociation Rates (PHIDRATES) website at http://phidrates.space.swri.edu. NRLMSISE-00 model atmospheres were generated using code downloaded from NASA Community Coordinated Modeling Center website at https://ccmc.gsfc.nasa.gov/pub/modelweb/atmospheric/msis/

EMBT thanks Drs. K. Greer and L. Andersson of LASP for their helpful discussions on elements of this study, as well as Dr. M. West of ROB who provided assistance with retrieving the SWAP data.

EMBT, FGE and MP contributions are funded by the MAVEN Mission, contract NNH10CC04C. EMBT also received funding from the PROBA2 Guest Investigator program.

LYRA and SWAP are projects of the Centre Spatial de Liege, the Physikalisch-Meteorologisches Observatorium Davos (for LYRA) and the Royal Observatory of Belgium funded by the Belgian Federal Science Policy Office (BELSPO) and, in the case of LYRA, by the Swiss Bundesamt für Bildung und Wissenschaft.

The work by M. Dominique is funded by the Belgian Federal Science Policy Office through the ESA-PRODEX program.